# piRank: A Probabilistic Intent Based Ranking Framework for Facebook Search

Zhen Liao

Meta Inc.

zliao@fb.com

## ABSTRACT

While numerous studies have been conducted in the literature exploring different types of machine learning approaches for search ranking, most of them are focused on specific pre-defined problems but only a few of them have studied the ranking framework which can be applied in a commercial search engine in a scalable way. In the meantime, existing ranking models are often optimized for normalized discounted cumulative gains (NDCG) or online click-through rate (CTR), and both types of machine learning models are built based on the assumption that high-quality training data can be easily obtained and well applied to unseen cases. In practice at Facebook search, we observed that our training data for ML models have certain issues. First, tail query intents are hardly covered in our human rating dataset. Second, search click logs are often noisy and hard to clean up due to various reasons. To address the above issues, in this paper, we propose a probabilistic intent based ranking framework (short for piRank), which can: 1) provide a scalable framework to address various ranking issues for different query intents in a divide-and-conquer way; 2) improve system development agility including iteration speed and system debuggability; 3) combine both machine learning and empirical-based algorithmic methods in a systematic way. We conducted extensive experiments and studies on top of Facebook search engine system and validated the effectiveness of this new ranking architecture.

## 1 Introduction

Commercial search engines (e.g., Google, Bing, YouTube, Baidu, Facebook) play an important role in providing information based on users' queries in their daily lives. Ranking algorithm is one of the core fundamental components to determine the final order of results users will see on the search result pages (SERPs).

Developing a good ranking algorithm is never easy and there are lots of existing studies on different types of ranking approaches. Among them, vector-space-model [1], language model [2, 3], TF-IDF [4], BM25[5] are well-known traditional algorithmic approaches to sort documents based on their relevance with queries.

With the rapid growth of machine learning (ML) communities, learning to rank [6] has attracted more and more attention as one of the hot research areas in recent decades. One advantage of learning to rank approaches is that they can leverage all existing algorithmic approaches as features and combine them together to achieve a better ranking performance. Typically, such approaches are supervised and depend on offline human-rating based datasets. Such datasets are usually constructed through collecting graded relevance labels (e.g., perfect, great, good, okay, bad) from human annotators, and the ML models are trained from training dataset and measured on test set in terms of well-known metrics like normalized discounted cumulative gains (NDCG) [7] and expected reciprocal rank (ERR) [8], etc. Numerous ML ranking algorithms are invented in the past decades (e.g., ListNet [9] and AdaRank [10]), and among them LambdaMart [11] is one of the state-of-art approaches which won first place in Yahoo's learning to rank challenge [12] in 2009.

Besides human-rating based relevance measurement, user engagements (e.g., click, time-spent) are also key metrics and implicit feedback for user satisfaction. Usually, such a dataset is logged from search engine product logging infrastructures. Most existing approaches use click through rate (CTR) as optimization metrics and build ML models based on the user click behaviors as positive labels. Different from human rating-based dataset (e.g., at most million level data samples), such dataset is typically at a much larger magnitude (e.g., million to billion level data samples), and thus more advanced ML approaches can be applied. Among them, Deep neural network-based approaches (DNN) [13, 14] are one typical type of state-of-the-art approaches.

Modern commercial search engines often leverage the knowledge obtained from both offline human-rated data and online user engagement data. In practice, we may combine the relevance prediction ML model trained from human-rating data and the click prediction ML model trained from user engagement data. However, as we improved the Facebook search system, we observed the following issues from such ML-based approaches due to their strong dependency on the training data.

On one hand, for offline human rating dataset, it is hard to scale. Due to the constraints of scalability, lots of query intents might not be fully explored or understood well through dynamic changing document inventories. In Facebook search logs, there is a non-trivial portion of tail query intents which are highly personalized and hardly been covered by our human rating datasets.

On the other hand, for online user engagement dataset, we found that it has a certain signal to noise ratio. First, the user clicks rely heavily on the current state of the system. If a perfect document of a query is never returned at the top of the SERP, we will never receive a click signal on it and thus will not rank it at top in the future. A hidden feedback loop such as this one is even harder to break when the click prediction ML model (i.e., click model) depends on memorization features (e.g., query-document pair click count or ratio, etc.), and the final ranking significantly depends on the model output. Second, there is a non-trivial volume of low-quality clicks (e.g., click-bait) where users are attracted by the document due to various reasons (e.g., attractive document thumbnails, title, abstract, etc.). Third, the query traffic is keeping changing from time to time, and thus we need to re-train our click models regularly from the latest training data. If a commercial search engine is actively being improved, then its search logs distribution will also keep changing. This will make the data cleaning work very challenging since the issues of the user engagement data may evolve as the system is updating dynamically.

Since ML-based approaches strongly depend on the quality and amount of training data, the above issues raise great challenges and block search systems to perform better in terms of ranking quality.

To address the above challenges, we propose a new probabilistic intent-based ranking framework named piRank (here **pi** is short for **p**robabilistic **i**ntent), that combines both ML-models and algorithmic approaches in a probabilistic way. This approach takes a few steps to model the composition of the final ranking score of a document given a query. First, a query from a specific user can have its own intent distribution. Second, each sub-intent can have its own ranking function which can be either ML-model or algorithmic or heuristic based. Third, those common factors for ranking will be extracted as common scoring components, and those intent-specific ranking logics will remain inside the sub-intent components. Finally, the ranking function consists of 2 parts: common ranking components and intent-specific ranking components.

The piRank approach has the following advantages compared to the existing approaches. First, it is scalable to allow adding new intent-specific ranking components. This can address the query intent shifting issues of a search engine and improve the performance of given sub-intent. Second, it can improve iteration speed and system debuggability in a commercial search engine with hundreds of engineers improving the ranking algorithms. For example, a "news" and a "entertainment" intent can develop their own intent-specific components in parallel. We can also easily drill down the ranking problem of a query to be common component issues or intent specific issues. Third, it can combine both algorithmic and ML based approaches and thus utilize the advantage from both sides. On the one hand, intent-specific rules can help to improve the quality of training data. The ML models can then be updated with better performance. On the other hand, the tail cases which are not addressed by ML-models can be identified and conquered by applying sub-intent specific ranking logic in a fast manner.

We tested the proposed framework at Facebook and up to now it has been applied into Facebook search production with significant offline and online evaluation metrics observed from A/B tests. We will present several launches in Section 5 to validate the effectiveness of piRank.

The rest of this paper is organized as follows. In Section 2 we will briefly introduce the related work. In Section 3, we will present the unique issues and status of Facebook search from both product and system perspectives. We will then describe our proposed intent-based ranking framework in detail in Section 4. In Section 5, we present the empirical study results through A/B testing at Facebook. We conclude this paper in Section 6.

## 2 Related Work

Existing work in search ranking algorithms mainly falls into two categories: unsupervised and supervised approaches. The unsupervised approaches are often algorithmic based and can be leveraged as input features to the supervised approaches. Among them, vector-space-model [1], language model [2, 3], TF-IDF [4] and BM25[5] are the most well-known and widely applied methods. Among them, language model is a probabilistic approach to model the likelihood of generating a query given a document in the term/word space. Different from it, the proposed approach in this paper models the query by different intents at a higher level.

Supervised learning to rank approach has attracted a lot of attention in recent years. Typically, such an approach requires training data which is annotated by human raters. Originated from the Yahoo challenge on learning to rank [12], LambdaMart [11] is one of the state-of-the-art methods. While methods like LambdaMart have been successfully applied to industry with great success, they heavily rely on the quality and the amount of training data and ranking features. For instance, we may assume that most of the relevant documents are rated per each query in our dataset. However, in real application scenarios, perfect documents might not be retrieved or ranked at the top for some queries, and therefore we might never rate them. On the other hand, we may assume that the query sampling is representative enough for the real query traffic. However, in practice, we may not be able to provide sufficient human annotated data which would cover different types of cases.

Besides building ML models from human rating dataset, another type of ML ranking approach is to model the user online behavior (e.g., click, time-spent, etc.). Typically, the click and non-click signals are treated as positive and negative labels and the model is trained from the huge amount of search logs. Traditionally, such click and skip information can also be modeled into pairwise labels and ML models (e.g., SVM-Rank [15]) can be applied on top of such constructed dataset. Nowadays, given the huge amount of training data and latest progress in various machine learning communities, we often apply models with huge capacity (e.g., deep neural networks) [13] to learn the personalized knowledge from the data to help ranking. While deep learning technologies from user engagement data can boost the performance of the system, it still has certain constraints and issues which block the future improvement of the system since user engagement behaviors are often biased towards the status of the system. First, a non-relevant document can also be clicked if it is ranked high. Certain studies [16] have been conducted to model the click-bias and examination process of the SERP. Second, a perfect document of a query which is never returned to top will not receive any click or user engagement. This also brings a cold-start problem to the training data. Third, the user engagement data keeps changing dynamically given both the improvements in search engine as well as user intent shift from time to time. This might force us to retrain our click models from the latest data occasionally. Thus, it is very hard to keep all historical knowledge we observed from the user engagement behavior. One practical solution is often to continuously update old model parameters based on new training data [17]. However, this won't work if we change the architecture or add new features to the model.

As a result, commercial search engines often combine human-rating based ML models and online click prediction models to balance the relevance and user engagement metrics in the system [18, 19]. In this paper, we propose a probabilistic intent-based approach to combine different types of algorithmic and ML ranking models together.

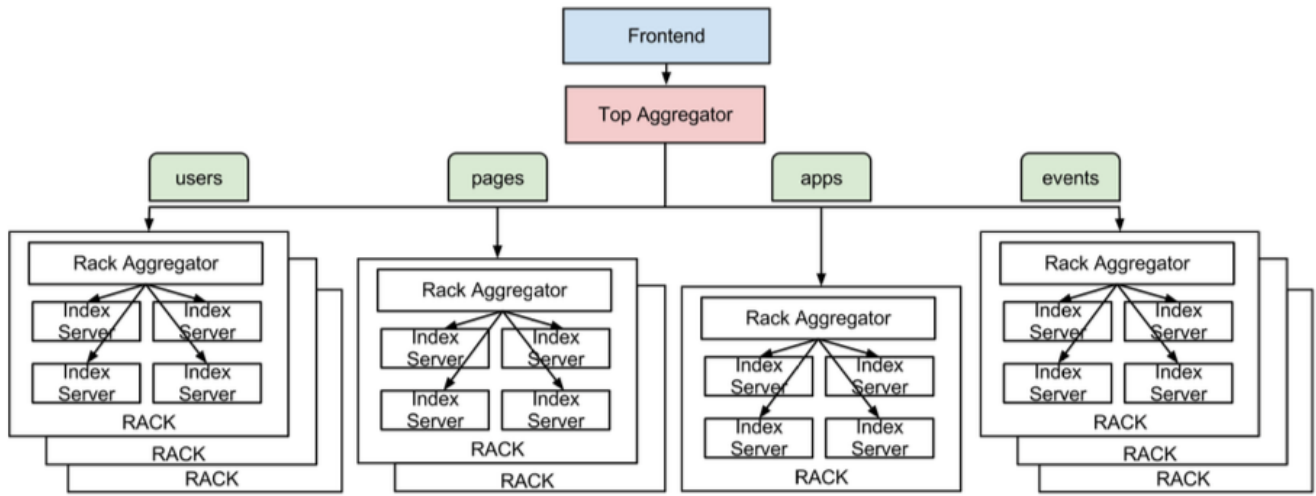

**Figure 1: The Facebook Unicorn System, picture from [20]**

## 3 Context of Facebook Search System

In this section, we will briefly introduce the background of this paper: the Facebook search product as well as its backend system.

Facebook search provides extensive information to users about their social networks (e.g., friends, pages, groups, etc.) as well as public content (e.g., public videos, posts, photos, etc.). From the Facebook app home screen, a user can start their search using the search icon. Taking query "Taylor Swift" as an example, when users type this query, we will return a blended list of documents (e.g., videos, pages, posts) by default in the "All" tab. Users can further click into sub-tabs like "Posts", "Videos", "Groups" to further filter on any specific types of information. We call the ranking of sub-tabs such as "Posts", "Videos", "Groups", etc. the vertical ranking, and the ranking of "All" tab as whole page ranking (WPR). One can test the search functions by visiting www.facebook.com or the Facebook app on cellphone.

Facebook search engine is built on top of a scalable search infrastructure named Unicorn [20]. Figure 1 illustrates the architecture of Unicorn with multiple search verticals at a high level. It provides index retrieval and ranking services for different types of documents (e.g., user-profiles, pages, posts, videos, events, etc.). When a user issues a query to the search engine, it will call the Top-Aggregator service. The Top-Aggregator will then send multiple sub-requests to different Rank Aggregators where each Rank Aggregator will fetch the documents from different index service nodes. At each index service node, each (query, document) pair will be scored and ranked. Then each Rank Aggregator will return the top results to the Top Aggregator. All documents returned to Top Aggregator will then be scored and sorted together. In this paper, to simplify the problem, we will mainly study the ranking function in the Top Aggregator.

## 4 piRank Approach

In this section, we will present the proposed approach in detail. We will start from describing the generic approach and then provide an instance based on the practice of Facebook search. We will also compare the proposed approach with other similar ML approaches like ensemble and mixture of expert models.

### 4.1 Generic piRank framework

Let q denote a query issued from a user. Note here q is personalized and contains personal contextual information from the user. Let d denote the candidate document which will be scored. Our goal is to find a ranking function $F(q,d)$ which can provide a score to sort all candidate documents. Please note that in this paper the proposed approach is focusing on $F(q,d)$, and we will not tackle problems like ranking diversification which can be done after the ranking order is determined.

Let r denote the ultimate relevance between a query q and document d. We predefine $F(q,d) = P(r|q,d)$ in the rest of this paper. Assume there is a query intent space T where each sub-intent $t \in T$ describes a specific query intent in the whole intent space. Note that there can be different query intent space (e.g., like different taxonomy of queries). For example, one instance of T can be {Sports Intent, News Intent, Music Intents, Movie Intents, Celebrity Intents, etc.} where each intent corresponds to a physical meaning of a query, and another instance of T can be {Video Intent, User Intent, Page Intent, Group Intent} where each intent corresponds to a specific object type of the documents. We will describe more specific instance in Section 4.2.

Assume each query sub-intent t is mutually exclusive to others, then based on the factor that $\sum_{t \in T} P(t) = 1$, we can have:

$$P(r|q,d) = \sum_{t \in T} P(r,t|q,d) \quad (1)$$
$$= \sum_{t \in T} P(r|t,q,d) \cdot P(t|q,d) \quad (2)$$
$$= \sum_{t \in T} P(r|t,q,d) \cdot P(t|q) \quad (3)$$
$$= \sum_{t \in T} F_t(q,d) \cdot P(t|q) \quad (4)$$

The formula (2) to (3) are based on independence between t and d. Basically a query intent is determined only based on the query and therefore we can ignore d. Here $F_t(q,d)$ is corresponding to the intent-specific ranking function.

Next, we assume that $F_t(q,d)$ is a linear combination of different ranking factors. Such linearity will improve the interpretation of the ranker, improve the debuggability of the system, and accelerate the iteration speed of the system.

$$F_t(q,d) \propto [\sum_c w_c \cdot \sigma_c(q,d)] + w_t \cdot \sigma_t(q,d) \quad (5)$$

Here c denotes the common ranking factors and s denotes the intent specific ranking factors. $w_c$ denotes the weight of factor c, and $\sigma_c(q,d)$ is the generic ranking model of c which compute the score, it can be either human rating based or user engagement-based ML models, or other algorithm approaches. Similarly, $w_t$ denotes the weight of intent specific factor, and $\sigma_t(q,d)$ denotes the ranking model which compute score of ranking factor s which is intent-t specific. Note that $\sigma_t(q,d)$ can be either ML-models from intent-t specific training data, or empirical rules built by domain expert or product specialist about intent-t.

Combing formula (5) and (4), we can get:

$$P(r|q,d)$$
$$\propto \sum_t P(t|q) \cdot \{\sum_c w_c \cdot \sigma_c(q,d) + w_t \cdot \sigma_t(q,d)\} \quad (6)$$
$$\propto \sum_c w_c \cdot \sigma_c(q,d) + \sum_t P(t|q) \cdot w_t \cdot \sigma_t(q,d) \quad (7)$$

Here the second part of (7) is the intent-specific ranking function, where for each intent it has three parts: $P(t|q)$ which can be viewed as the intent score to decide the triggering of this intent score component, $w_t$ as the weight to balance different intent's weight, and $\sigma_t(q,d)$ as intent-specific relevance scoring function. Note that both $w_c$ and $w_t$ can be pre-selected parameters or output scores from other ML models.

### 4.2 Instances of piRank

Based on Formula (7), a ranking function can be separated as generic ranking components and intent-specific ranking components. In the following, we will show possible examples of both $\sigma_c(q,d)$ and $\sigma_t(q,d)$.

*4.2.1 Generic ranking components.* The $\sigma_c(q,d)$ can be those ranking algorithms which are important among different query intents. We will present a selective list of such functions in the following section.

**Text Relevance**. Text relevance can be a ML or algorithmic function based on TF-IDF, BM25, Text Proximity, and word

embedding based text similarity between a query and a document. To combine all such features in a systematic way, ML approaches based on human rated data can be applied here. More specifically, a dataset based on graded relevance (e.g., perfect, great, good, okay, bad) can be constructed with human rating and then one can train a LambdaMart model to optimize for the NDCG metric on the evaluation dataset. Alternatively, one can also train a regression decision tree model (e.g., Gradient Boosting Decision Tree [22]) to predict the relevance grade of a query-document pair and use that as a ranking component.

**User Engagement.** The user engagement component is another ranking component to improve the ranking performance. Usually, such functions will be a ML model trained on large scale user engagement data logged from search engines to predict the probability of engagement of a new (user, query, document) tuple. The label of such data can be click, click with follow-up actions (e.g., click a user profile and make a conversation with the user, click a post, and leave a comment), click with certain dwell time (e.g., click a video and spend more than 1 minute watching the video), or tail engagement events (e.g., like/comment/share of a post on the SERP). Important features of such engagement models are historical engagement information since such engagement histories are often labels of the previous dataset, and users would like to engage with similar content they have engaged before. Besides the historical engagement information, categorical features like query intent category, pairwise features like query-document matching, and certain sparse features are also important. In practice, one can apply the deep neural network model [21, 23] and an example architecture are presented in Figure 2.

**Social Relevance.** The social relevance is another horizontal and generic ranking relevance factor to be considered at Facebook. We can consider all users and documents on Facebook as a social network and extract the possible relationship between any two nodes (e.g., a note can be either a user or a document). For instance, a feature can be a relationship between the searcher (user) and the author of the candidate document. Other examples of such relationships might be self (the document is published by the searcher), friend (the document is published by a friend of the searcher), friend-of-friend (the document is published by a friend of a friend of the searcher), self-engaged (e.g., the document has been clicked by the searcher) friend-engaged (e.g., the document has been clicked by a friend of the searcher), followee (e.g., the document is a page or person followed by the searcher), follower (e.g., the document is a person who followed the searcher), pending-friend (e.g., the document is a user who sent a friend request to the searcher), pending-joining (e.g., the document is a group who has a pending join request from the searcher), etc. These social relevance features are important especially for users who want to revisit the information they've seen before on Facebook or check updates or opinions from their social networks. We can combine such social engagement information as features into the user-engagement-model described above or train a social-engagement specific ML model from engagement information on socially connected documents or combine such features together in algorithm way (e.g., either linear or non-linear into the final ranking function).

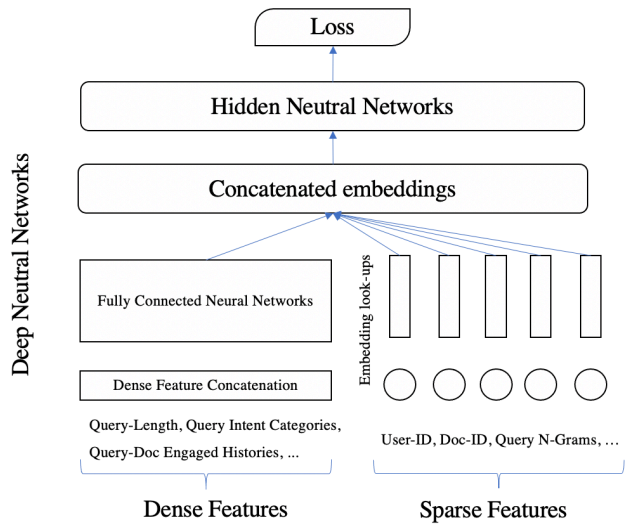

**Figure 2: An Example of Deep Neural Network Architecture of User Engagement model.**

**Location Relevance.** Location information is another important contextual information of a searcher, which can be used to help with lots of different types of query intents. For instance, query "coffee" might have an implicit intent to find Starbucks or other coffee shops nearby, query "new mom groups" might have the intent to explore mom groups in nearby areas, query "weather" may mean looking for weather information of the searcher's city, etc. The location relevance can be a ranking function built on top of location features. Such features can be (user, document) distances measured by latitude/longitude, or (user, author of the document) location distances, or the average distance of (user, engaged users of the document), etc.

**Language Matching**. Language matching is another important factor to consider in a global search engine serving document from different languages. For instances, by searching "taylor swift" one can get a music video with Arabic language while the video description has the text "taylor swift". In order to compute the correct language matching features, we compute the matching probability of (user, document) pair as well as (query, document) pair. For a user, we infer their language from the posts they've published or shared as well as the pre-defined language in the profile. For a document, we infer its language from the author of the document, the engaged user of the document. We have also built language classifiers which can help us in classifying the text language of a document among hundreds of different languages on Facebook. Such classifiers can be used to classify the language of a document or a query. For specific documents such as videos with very sparse textual information, we have also built audio classifiers which can classify the language of the audio in a video among hundreds of languages with a very high precision and recall. All such language matching information is combined as a language matching component in the generic ranking function.

**Document Quality.** Different from other generic ranking components which need both query and document side information, document quality only depends on the document itself and thus can be pre-computed before the search happens. For instance, a document quality feature might be kids-friendly (e.g., if a document contains any adult-only content like porn/sexual information), authentic (e.g., whether a document contains fake information), authoritative (e.g., if the document is published by an authoritative source or a random source), readiness (e.g., whether the textual information of the document is readable and easy to understand), video resolution (e.g., whether a video has good resolution to be watched), etc. Due to company policies, some documents will be rejected by the ranking layer if a certain dimension of the document quality has issues (e.g., a porn video will be filtered).

There might be other types of generic ranking functions which work among different query intents, we selectively list and present several of them which have been tested as effective and are among the most important ones in our system.

*4.2.2 Intent-specific ranking components.* While the generic ranking components of $\sigma_c(q, d)$ can be useful to improve the ranking performance in horizontal way, we may also need to have intent-specific $\sigma_t(q, d)$ which can further boost the performance of the search. In the following we will present several examples about intent-specific ranking components. Certain components are typically validated from our user studies as well as A/B tests. We will describe the intent detection and user expectation on specific intent in Section 4.3.

**Friend intent.** Friend intent means the searcher is searching for their friends. This is a very special use-case of Facebook search since a huge number of users are connected on Facebook. For such intent, the intent specific function can be matching the target friend's profile, or matching the publisher of the videos / posts / photos by the friend, etc. It can also be a post that is reshared or commented by the friend or a post tags the target friend.

**Special Grammar intent**. Special grammar intent is a common usage of Facebook search where users typed "posts I have seen" or "videos I watched yesterday" and try to find the desire results. For such specific intent, the intent specific scoring function needs to consider specific grammar type to matching with desire intent. Such grammar matching will be highly personalized.

**Video publisher intent**. There are certain queries searching that seek videos from Facebook publishers such as "5-minutes crafts". This is a specific intent, and it requires specific video publisher (e.g., specific entity) matching. Therefore, the

$\sigma_t(q, d)$ can be a binary function for publisher matching or a ML model which indicate the matching confidence between the query and the video publisher.

Above intent specific ranking components are selected based on their prevalence at Facebook. There can be a long list of tail intents which are not covered in the list above. While building the intent-specific ranking component, we recommend considering the query traffic of certain intent to make a good trade-off. It is also worth noting that not all queries can be categorized into a clear query intent, and such cases can be supported by the general ranking components.

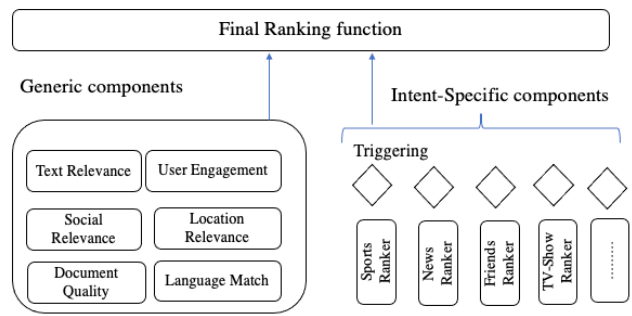

**Figure 3: An instance of piRank architecture.**

*4.2.3 piRank architecture at Facebook.* By putting all elements together, the entire piRank architecture is presented in Figure 3. Please note that since Facebook search has multiple search verticals (e.g., users search, pages search, videos search, posts search, etc.), so each search vertical can have their own piRank and the entire whole page ranking can also apply the piRank framework. In practice, this architecture is also easy to debug and monitor. For example, for a given specific intent of queries, we can easily check why an ideal document is not ranked high by checking the weights and scores of generic and intent-specific ranking components. Another benefit of intent-specific ranking component is that we can keep iterating on different types of ideas through A/B testing and use the final user engagement feedback to validate different hypotheses. Through a real-time logging of intent-triggering as well as ranking scores changes, we can also build alerts to report issues if certain intent-specific components suddenly get triggered more than usual to help detect system bugs or query intent distribution changes. Therefore, the proposed piRank architecture can improve the system development iteration speed and make the entire ranking function easy to debug, monitor and understand.

## 4.3 Intent Detection and User Expectation

For the piRank framework, one important component is to figure out the query intent. To do that, we have built a comprehensive intent detection framework integrated into Facebook Search System. The intent detection system has a few different components listed below.

**Entry Point Disambiguation**: where we use the clicked query suggestion with structured information (e.g., with the friend profile, etc.) to understand the intent of the query. This step was done by adding rich information to our typeahead system (e.g., auto-completion). When user input a prefix of a query, we will provide several auto-completion query suggestions. Such query suggestions are different from previous plain text suggestion, where we will provide rich information like user profile pictures, page profile pictures, rich snippets to help understand the intent of the searcher. For example, when user typed "tom" in search bar, we will provide a few suggestions like: {1: tom cruise, actor & director, picture-1}, {2: tom hanks, actor, picture-2}, {3: tom and jerry, movie of 2021, picture-3}. When users clicked into any of the suggestion, it will help to disambiguate the input and make it a clear intent query. From our current production traffic, around 20% of traffic falls into this category with entry point disambiguation.

**Query Intent Classification**, where we build hundreds of classifiers for a variety of query intents (e.g., news, sports, entertainment, etc.). The intent classifiers have two major themes. One theme of classifiers is neutral network-based model, which are built from offline rating data with binary labels and use the character level input from queries. Such classifiers are built from general intents and commonly used labeling data. On average, the F1 score (the harmonic mean of the precision and recall) of such classifiers are around 0.7 based on our offline evaluation. The second theme of classifiers is built from different engineering team based on the needs of their domains. For example, our people ranking team built specific people classifier to better identify the people search intents. Such classifiers can take more advanced features (e.g., user's social graph information) into consideration and performing better than general intent models. On average, the F1 score of such classifiers are around 0.8 based on our offline evaluation. The query intent classifiers can classify around 60% of total traffic with high confidence (e.g., with precision higher than 80%).

**Entity Linking:** here the problem is to identify the target entity ID from the input query. For example, given a query "Adele hello" we should identify the page ID of "Adele", and the song ID of "hello" in our database. To make this even more complex, we also need to provide the entity-ID like users' friends or groups or friends of friends. Since there are lots of duplicated names and entities of the same short text, we leveraged deep contextual modeling [17] to help detect the correct entities inside the query. The basic idea is to leverage the contextual words of the entity mention inside the query to improve the accuracy. On average, the F1 score of entity linking is around 0.6 based on our offline evaluation given the complexity of the problem. The entity linking can help to identify around 50% of query traffic with relatively high confidence (e.g., precision higher than 70%).

**Query Pattern**: query patterns are built on top of entity linking or other predefined rules. For example, we can define a rule like {<movie:entity> trailers} which can match to queries like {*avengers* trailers}, {*life of pi* trailers}, where the movie names can be detected. We can also build a dictionary for trailers pattern using plain list of keywords. Then the rule can be modified as {<movie:entity> <trailers:dictionary>}. Then it can be used to matching with different keywords for trailers (e.g., using different languages). We have built ~600 such patterns to covering a wide range of queries (e.g., ~30% of query traffic). The precision of such patterns is based on either our regex rules or entity linking. Most of such patterns are above 90% accurate and can be used as good rules to enhance the ranking of the system.

With our intent detection system, we also run user survey and research to understand the most desired documents for specific intents. Through collecting around ~190,000 of user survey feedback across ~20 countries, we summarize and get the top user expectations of different query intents and describe them in the Section 4.2.2.

Note that there might be still ~10% of query volume which are tail or ambiguous and can hardly be covered by any intent detection components with high confidence, and we can just rely on the general ML models to rank documents for them.

## 4.4 Weights of piRank: PE-BVTs

As described above, piRank introduces parameters of weights for both generic and intent-specific ranking components in a linear combination way. Such weights can bring advantages like easy to monitor, debug and interpret. We may need to bring a strategy to adjust the weights when we introduce new components or make changes to the overall ranking function (e.g., adding new features to a ML model, revise intent-specific ranking logics, etc.). In practice, when we make simple changes (e.g., adding a new intent-specific component, slightly improve generic ranking functions), one can slightly tune the parameter of corresponding component through conducting A/B testing on a range of weight values. At the meantime, we also developed a systematic quality evaluation framework which is intent-based product expectation basic verification tests (PE-BVTs), to help monitor the quality and relevance for different query intent, to serve as both quality guardrails and debugging tool for ranker tuning.

The PE-BVTs are developed in a way where given the input query and user context, we expect the search system to return certain documents which can satisfy the pre-defined expectations. For example, taking friends search as an example, if a user uses her friend's name to do a search, we expect the system to return the corresponding friend profile as top-1 result. We totally developed ~200 PE-BVTs to covered ~80% of Facebook search traffic and number of PE-BVTs is still increasing to cover more cases.

The advantages of PE-BVTs are: 1) it can help to understand the corresponding traffic of certain query intent; 2) it can scale up to different query languages; 3) it does not require human rating efforts; 4) it can be good debugging tool to help root-causing the failures of a search; 5) it can avoid search relevance regression; 6) it can also help big system refactoring to avoid unexpected system behaviors; 7) it is sensitive to query intents which have small traffic.

## 5 Empirical Studies at Facebook

In this section, we will present the empirical study at Facebook for the proposed solution. Due to company policy, we will report the delta of metric gains instead of absolute values.

We will present different types of A/B testing experiments by adding intent specific components, adding generic ranking components, updating ranking components, as well as weight-tuning. All such experiments can validate the effectiveness and flexibility of the proposed piRank framework in commercial search engine. We will use video search to showcase the effectiveness of this approach and briefly discuss the application for other search applications. From our past practice, we have observed similar effect by applying piRank to other search verticals as well.

The key metrics we used to track in the A/B experiments are SERP good click rate (SGCR), which is defined as user's click on a SERP as well as certain good interaction after the click (e.g., spend certain times, make a comment, etc.). This metric has been studied by checking its correlation with other logging metrics (e.g., search volume, reformulation rate, etc.) as well as user survey (e.g., the SGCR and user survey-based satisfaction have high correlation). We can view SGCR as a variant of SCR (SERP click rate) but it can help denoise the low quality part of SCR. In the following experiments, the improvement of SGCR is statistically significant (with p-value < 0.05).

### 5.1 Intent Specific Components

The first A/B test experiment was in the context of a video search product, where we observed lots of queries with clear video publisher intent and did not get the corresponding results at top. A typical example of such type is "5 minutes craft" which is a popular video publisher name. Without the intent specific boosting the SERP has lots of low-quality videos with the text matching. Through this boosting, the SGCR of the test arm increased 0.43% (with p-value < 0.01). Also, the total time-spent on watching videos of the test arm is 0.83% higher than the control arm. Here the intent specific component is just a simple video publisher matching.

After launching the test-arm into production for a few weeks, we updated the intent-specific component to be video publisher matching as well as good-clicks. Through this we were able to prevent boosting videos from "taylor swift" official page started to dominate the SERP. The follow up experiment brings improved SGCR by 1.5% compared to the baseline with just binary intent-specific ranking indicators.

The video publisher intent component shows an example of adding intent-specific components to adjust the ranking based on existing generic ranking components.

### 5.2 Generic Components

Besides the intent-specific components, below is an example of adding generic ranking components into the ranking function. In videos search we observed videos with languages which are not known by the end users. This is due to lots of videos containing both the query and some other languages. To fix this, we introduced a language matching generic component which computes the matching score between user and document by using the technology described in Section 4.2.1. Through this, it improved SGCR by 0.46%.

### 5.3 Components Updates

As the system keeps changing, the top ranked documents for a query also change. Therefore, it is helpful to use the latest training data which reflects the latest status of the system. Both human rating and online user engagement data will help capture the latest issues of the system. Here we run a test of updating the generic ranking components by retraining the relevance model using the latest batch of human rating data. Through this updating, we observed the score distribution shift, so slightly adjusted the weight of the model. The experiment brought about 0.07% SGCR gains compared to the old ranker.

### 5.4 Weight Tuning

After we introduced multiple intent specific components and generic ranking components, the overall ranking function now has around 20 ranking components. At this stage, tuning weights of all ranking components together manually is not practical. Therefore, we applied a heuristic parameter search approach based on offline PE-BVTs evaluation, which can help to automatically tune and find the best parameter set based on the metric of SGCR. From the experiment, we improved SGCR by 1.23%.

### 5.4 piRank For Other Applications

We have also launched piRank to other search verticals including people, groups, page, photos, and posts ranking, as well as whole SERP (search engine result page) ranking with heterogeneous objects. For different verticals, the piRank components as well as their weights are customized to achieve the best performance of the corresponding vertical. For example, for people ranking we have general components to predict the click and friending behaviors (e.g., how likely will the searcher send a friend request the result user as friend), and intent specific components for friends and friends-of-friends. For groups ranking, we have general components to predict the click and join behaviors (e.g., how likely will the searcher join the returned group on the SERP), and intent specific components for special topics (e.g., location aware groups or sensitive topics including vaccination etc.). For whole SERP ranking, we have general components to predict the click and relevance, and intent specific components to rank special types of objects (e.g., video specific intent, groups specific intent, friend specific intent, page specific intent, etc.).

## 6 Conclusion

In this paper, we proposed a probabilistic intent based ranking framework named piRank which takes a probabilistic approach to combine generic ranking components and intent-specific ranking components. The proposed approach can address the issues where generic ranking components do not work well due to lack of high-quality training data, and thus it can help to improve the search system into a better stage and improve the quality of the training data to all ML ranking models. It has been widely used in Facebook search system with many teams of engineers to help parallelize the development of ranking system improvements and saved the debug cost from finding root-cause of a sub-par or counter-intuitive ranking results. We have productionized the proposed approach in a commercial search engine in the past few years to validate its effectiveness.

## ACKNOWLEDGMENTS

The author would like to thank Abhinav Sangal and Hailing Cheng for proof-reading and helping provide changes and suggestions for the paper, thanks also to Feilong Liu, Shenggu Lu, Pramodh Karanth Prabhakar, Chiyao Shen, Mayank Singh, Sadegh Mirshekarian for implementing the algorithms and experimentation. Thanks also to Nikhil Garg, Pei Yin, Guangdeng Liao, Bi Xue, Yinzhe Yu, and Hong Yan for supporting the projects and insightful discussion about the work. There are lots of other great collaboration with Meta colleagues who have been involved in the development, discussion and experimenting with the proposed approach in this paper.

## REFERENCES

[1] Gerard Salton, Anita Wong and Chung-Shu Yang, 1975. A Vector Space Model for Automatic Indexing. Communications of the ACM 18, 11, 613–620.
[2] Jay M. Ponte and W. Bruce Croft, 1998. A language modeling approach to information retrieval. 1998. In Proceedings of the 21st annual international ACM SIGIR Conference, 1998.
[3] Fei Song and W. Bruce Croft. 1999. A General Language Model for Information Retrieval. In Proceedings of the eighth international conference on Information and knowledge management. Nov. 1999 Pages 316–321
[4] Spärck Jones, K. (1972). A Statistical Interpretation of Term Specificity and Its Application in Retrieval. In Journal of Documentation. 28: 11–21. doi:10.1108/eb026526.
[5] Stephen Robertson and Hugo Zaragoza (2009). The Probabilistic Relevance Framework: BM25 and Beyond. Foundations and Trends in Information Retrieval. 3 (4): 333–389. doi:10.1561/1500000019
[6] Tie-Yan Liu, 2009. Learning to Rank for Information Retrieval. Foundations and Trends in Information Retrieval. March 2009
[7] Kalervo Järvelin, Jaana Kekäläinen: Cumulated gain-based evaluation of IR techniques. ACM Transactions on Information Systems, 422–446 (2002)
[8] Olivier Chapelle, Donald Metlzer, Ya Zhang, and Pierre Grinspan, 2009. Expected reciprocal rank for graded relevance. In Proceedings of the 18th ACM conference on Information and knowledge management. November 2009 Pages 621–630 https://doi.org/10.1145/1645953.1646033
[9] Zhe Cao, Tao Qin, Tie-Yan Liu, Ming-Feng Tsai and Hang Li. Learning to rank: from pairwise approach to listwise approach. In Proceedings of the 24th international conference on Machine learning. June 2007.
[10] Jun Xu and Hang Li, 2007. AdaRank: a boosting algorithm for information retrieval. In Proceedings of the 30th annual international ACM SIGIR conference. July 2007. Pages 391–398.
[11] Christopher J.C. Burges, 2010. From RankNet to LambdaRank to LambdaMART: An Overview. Microsoft Research Technical Report MSR-TR-2010-82.
[12] Olivier Chapelle and Yi Chang, 2011. Yahoo! Learning to Rank Challenge Overview. In Proceedings of Machine Learning Research. 2011
[13] Paul Covington, Jay Adams, Emre Sargin. 2016. Deep Neural Networks for YouTube Recommendations. In Proceedings of the 10th ACM Conference on Recommender Systems. September 2016 Pages 191–198.
[14] Guorui Zhou, Xiaoqiang Zhu, Chengru Song, Ying Fan, Han Zhu, Xiao Ma, Yanghui Yan, Junqi Jin, Han Li and Kun Gai. 2018. Deep Interest Network for Click-Through Rate Prediction. In Proceedings of the 24th ACM SIGKDD International Conference on Knowledge Discovery & Data Mining. July 2018 Pages 1059–1068.
[15] Thorsten Joachims, 2002. Optimizing Search Engines Using Clickthrough Data, Proceedings of the ACM Conference on Knowledge Discovery and Data Mining (KDD), ACM, 2002.
[16] Nick Craswell, Onno Zoeter, Michael J Taylor and Bill Ramsey, 2008. An experimental comparison of click position-bias models. In Proceedings of the 2008 International Conference on Web Search and Data Mining. February 2008 Pages 87–94.
[17] Zhen Liao, Xinying Song, Yelong Shen, Saekoo Lee, Jianfeng Gao and Ciya Liao, 2017. Deep Context Modeling for Web Query Entity Disambiguation. In Proceedings of the 2017 ACM on Conference on Information and Knowledge Management. November 2017. Pages 1757–1765.
[18] Dawei Yin, Yuening Hu, Jiliang Tang, Tim Daly, Mianwei Zhou, Hua Ouyang, Jianhui Chen, Changsung Kang, Hongbo Deng and Chikashi Nobata, Ranking Relevance in Yahoo Search. In Proceedings of the 22nd ACM SIGKDD August 2016 Pages 323–332
[19] David Alexander Sontag, Kevyn Collins-Thompson, Paul N Bennett, Ryen W White, Susan T. Dumais, Bodo Billerbeck, 2012. Probabilistic models for personalizing web search. In Proceedings of the fifth ACM international conference on Web search and data mining. February 2012, Pages 433–442 https://doi.org/10.1145/2124295.2124348
[20] Michael Curtiss, Iain Becker, Tudor Bosman, Sergey Doroshenko, Lucian Grijincu, Tom Jackson, Sandhya Kunnatur, Soren B Lassen, Philip Pronin, etc. 2013. Unicorn: a system for searching the social graph. In Proceedings of the VLDB Endowment. August 2013.
[21] Hengtze Cheng, Levent Koc, Jeremiah J Harmsen, Tal Shaked, Tushar D Chandra, Hrishikesh Balkrishna, etc. 2016. Wide & Deep Learning for Recommender Systems. In Proceedings of the 1st Workshop on Deep Learning for Recommender Systems. September 2016 Pages 7–10
[22] Jerome H Friedman, 2002. Stochastic gradient boosting. In Computational Statistics & Data Analysis. February 2002.
[23] Masoudnia, Saeed; Ebrahimpour, Reza (12 May 2012). "Mixture of experts: a literature survey". Artificial Intelligence Review. 42 (2): 275–293. doi:10.1007/s10462-012-9338-y. S2CID 3185688.